\documentstyle[prl,aps,manuscript]{revtex}

\begin{document}
\draft

\title{Comment on ``Signal of Quark Deconfinement in the Timing
Structure of Pulsar Spin-Down''}
\maketitle

Glendenning, Pei, and Weber (GPW) [1] 
studied the possible 
observational consequences of the continuous quark-hadron phase 
transition in the core of a neutron star during its spin-down 
process. They concluded
that the star should contract 
from $r{\sim}15$ km to $r{\sim}10$ km as
it spins down from a Keplerian velocity to a non-rotating state. 
The spin-down behavior
would thus be dramatically different than that of a neutron star
without a phase transition. In addition to the ``braking index'',  
the post-glitch behavior may also be
affected by the increase in the total amount of quark matter
inside the star [2]. 
%In this Comment we show that 
%GPW's estimate of the event rate is incorrect, and 
%the gravitational energy release in the star, which is omitted in the
%calculations in [1], may lead to more important 
The event rate is very important for observational confirmation
of the signature of quark deconfinement. 
GPW estimate [1] that 1\% of the 
observed pulsars are in the process of phase transition 
on a time scale of $10^5$ years. 
This is based on simply assuming that {\it all} pulsars have the 
parameters favoring the phase transition. As we have 
shown [3], less than 0.1\% of all neutron
stars might undergo a phase 
transition in their lifetime. The 
reason is that the increase in central density 
due to spin-down is small 
(${\Delta}{\rho}_{\rm c}/{\rho}_{\rm c}{\sim}10^{-3}$), 
leaving  a very narrow window of densities favoring a phase
transition. For example, stars with central densities lower
than ${\rho}_{\rm crit}-{\Delta}{\rho}_{\rm c}$ can not reach the critical
density, and stars with initial central densities higher than the 
critical density should be born to have large quark cores.  
Hence, the 
possibility of finding this phenomenon among the $\sim$700 
radio pulsars is low. 

The beaming of pulsar radio emission 
prevents us from seeing the majority 
of neutron stars via radio 
observations. GPW neglected the fact that
the contraction of the star is associated with the 
release of a huge amount of 
gravitational energy. The star heats up and radiates 
isotropically in X-rays, thus it is more promising 
to search for the signal of quark deconfinement among X-ray 
sources than among radio pulsars. 
The released gravitational energy can be estimated as
\begin{equation}
 \displaystyle
    E \sim \frac{GM^2}{R}\left(\frac{{\Delta}R}{R}\right) \simeq
10^{53} ~ \left(\frac{{\Delta}R}{R}\right) ~ {\rm ergs},  
\end{equation}
thus the average rate is $10^{40}$ ergs s$^{-1}$ in $10^5$ years. 
Van Riper has studied [4] the thermal response of a neutron
star to a steady state energy source, and finds that
for a central energy source of $10^{40}$ ergs s$^{-1}$, the star 
is heated up to a surface temperature of $3{\times}10^{6}$ K, 
yielding a soft X-ray luminosity of ${\sim}10^{35}$ ergs s$^{-1}$. 
The majority of the gravitational 
energy is released via neutrino emission. 

Among the ${\sim}10^8$ neutron stars in the Galaxy (the 
neutron star birth rate is about $10^{-2}$ per year per galaxy, 
and the age of our Galaxy is $10^{10}$ years), 
${\sim}10^5$ may have phase 
transitions. If each transition lasts $10^5$ years, 
we expect to see a few (on the order of unity) isolated 
bright ($\sim$20 times more luminous than the Sun) X-ray
sources without optical counterparts. 
Neutron stars with very strong magnetic fields 
(${\sim}10^{15}$ G)
have been suggested as candidates for the observed 
``soft $\gamma$-ray repeaters'' [5]. 
These are rare (only 4 observed) X-ray transient 
sources associated with young (${\sim}10^4$ years) supernova 
remnants, and usually are also quiescent X-ray emitters 
($L_{\rm X}{\sim}10^{35}$ ergs s$^{-1}$). I suggest that 
soft $\gamma$-ray repeaters may be related to neutron stars
with phase transitions. Neutron stars with strong magnetic fields
spin down faster than those with weaker fields, and thus
are more likely to undergo phase transitions. 
This conclusion is supported by the agreement of the 
predicted X-ray luminosity associated with phase transitions 
and that observed in  soft $\gamma$-ray repeaters. The predicted
and the observed event rates are also in agreement.  

The longer time scale 
%($10^7$ years for millisecond pulsars [6])
of the phase transition 
does not lower the X-ray luminosity significantly
because the cooling is 
dominated by neutrino emission once the central energy
source is larger than $10^{35}$ ergs s$^{-1}$ [4]. 
On the other hand, if the time scale is much shorter and is
comparable with neutrino diffusion time, the star
heats up to a much higher temperature. 
Before the realization by 
Glendenning [6] that there
are more than one conserved charges in the quark-hadron
phase transitions in neutron stars, models predict
a density discontinuity between the two phases. If this were 
the case, a phase transition happening during a neutron star's
spin-down process would be catastrophic (on a time scale
of seconds) and may power a $\gamma$-ray burst [3]. 

I thank Nicholas Matlis and David Chappell for help with the manuscript. 

%Finally I address that the spin-up of an isolated 
%millisecond pulsar due to quark deconfinement [6] 
%is actually powered by the release of gravitational
%energy. Detailed calculations including heating process
%and differential rotations are needed to study the 
%rotational behavior of pulsars undergoing phase transitions. 

\begin{flushleft}
Feng Ma\\
McDonald Observatory, University of Texas\\
Austin, TX 78712--1083\\
\end{flushleft}

\end{document}